\documentclass[a4paper,showpacs,twocolumn,amsmath,amssymb,floatfix,prb,preprintnumbers,footinbib]{revtex4}
\usepackage[dvips]{graphicx}
\usepackage{dcolumn}
\usepackage{bm}

\begin{document}

\newcommand{\comment}[1]{}
\newcommand{\T}{\mathcal{T}}


\title{Rashba effect in quantum networks} 

\author{Dario Bercioux$^{1,2}$, Michele Governale$^3$, Vittorio
Cataudella$^2$, and Vincenzo Marigliano Ramaglia$^2$}
\affiliation{$^1$Institut f{\"u}r Theoretische Physik, Universit{\"a}t
Regensburg, D-93040, Germany\\ $^2$Coherentia-INFM \& Dipartimento di
Scienze Fisiche Universit\`a degli Studi ``Federico II'', I-80126
Napoli, Italy \\ $^3$NEST-INFM \& Scuola Normale Superiore, Piazza dei
Cavalieri 7, I-56126 Pisa, Italy}

\date{\today}

\begin{abstract}

We present a formalism to study quantum networks made up by
single-channel quantum wires in the presence of Rashba spin-orbit
coupling and magnetic field.  In particular, linear transport through
one-dimensional and two-dimensional finite-size networks is studied by
means of the scattering formalism.  In some particular quantum
networks, the action of the magnetic field or of the Rashba spin-orbit
coupling induces localization of the electron wave function.  This
phenomenon, which relies on both the quantum-mechanical interference
and the geometry of the network, is manifested through the suppression
of the conductance for specific values of the spin-orbit-coupling
strength or of the magnetic field.  Furthermore, the interplay of the
Aharonov-Bohm phases and of the non-Abelian phases, introduced by
spin-orbit coupling, is discussed in a number of cases.

\end{abstract} 

\maketitle

\section{Introduction}
In recent years, a effect of extreme localization induced by
magnetic field has been predicted in a class of two-dimensional
rhombus tilings~\cite{vidal-1998}.  This effect is related to the
interplay between the Aharonov-Bohm (AB) effect~\cite{bohm} and the
geometry of the network. For special values of the perpendicular
magnetic field, the set of sites visited by an initially localized
wave packet is bounded by the AB destructive interference. This set of
sites is referred to as \emph{AB cage}.  This kind of localization
does not requires the presence of disorder~\cite{anderson-1958}, as it
relies on quantum-interference and on the geometry of the lattice.
There have been several theoretical works addressing different aspects
of AB cages, as the effect of disorder and electron-electron
interaction~\cite{vidal-prb-long}, interaction induced
delocalization~\cite{vidal-cage-2000}, 
transport~\cite{vidal-tra-2000}, and realizations  
in fully frustrated
superconducting systems~\cite{korshunov,cataudella}.

Two series of experiments have confirmed the existence of the AB-cage
effect. Abilio \textit{et al.}~\cite{abilio-1999} have shown that a
$\T_{3}$ network realized with superconducting wires exhibits a
striking reduction of the critical current and of the superconducting
transition temperature for the predicted values of the magnetic field.
Naud \textit{et al.}~\cite{naud-2001} have measured megnetoresistance
oscillations of a normal $\mathcal{T}_{3}$ network, tailored in a high
mobility two-dimensional electron gas (2DEG).

It is known that a wavefunction on an electron moving in the presence
of Spin-Orbit (SO) coupling acquires quantum phases due to the
Aharonov-Casher effect~\cite{casher,mathur,
balatsky,aronov-1993,splettstoesser-2003,frustaglia}.  We focus on the
Rashba SO coupling \cite{rashba-1960,rashba-1984}, which is present in
semiconductor heterostructures due to lack of inversion symmetry in
growth direction.  It is usually important in small-gap
zinc--blende--type semiconductors, and its strength can be tuned by
external gate voltages. This has been demonstrated experimentally
measuring the Shubnikov-de Haas oscillations~ 
\cite{nitta-1997,
schaepers-1998,grundler-2000,schaepers-2004} or
antiweak localization~\cite{miller-2003} 
in two-dimensional electron gases.

In a recent Letter~\cite{bercioux-2004}, it has been shown that in a
linear chain of square loops connected at one vertex (termed
\textit{diamond chain}), localization of the electron wavefunction can
be obtained by means of the Rashba effect.

In this paper the formalism introduced in our previous
work~\cite{bercioux-2004} is extended to include both Rashba SO
coupling and magnetic field. In this case, electrons traveling in the
network acquire both the AB phase factors and the non-Abelian phases
induced by SO coupling.  The main aim of the present paper is to study
the interplay of the magnetic field and of the SO coupling, and to
investigate two-dimensional networks with Rashba SO coupling
(previously only one-dimensional networks were studied).

The paper is organized in the following way. In Sec.~\ref{model} we
introduce the model and the formalism used to study a quantum network
realized with single-channel quantum wires in the presence of Rashba
SO coupling and of an external magnetic field. Section~\ref{one-dim}
is devoted to the transport properties of one-dimensional networks in
the presence of magnetic field and Rashba SO coupling. In
Sec.~\ref{twodim} results for transport through two-dimensional
networks are presented. Conclusions are drawn at the end of the paper.


\section{Model and formalism\label{model}}

We consider a two-dimensional network (in the $xy$ plane) made up of
single-channel quantum wires with Rashba SO coupling, and we allow for
the presence of a magnetic field perpendicular to the plane of the
network.

We start from the Hamiltonian of a single-channel wire along a
generic direction $\hat{\gamma}$ in the $xy$ plane:
%
\begin{equation}\label{full-hamiltonian}
        \mathcal{H}=\frac{\left(p_\gamma + q A_\gamma \right)^2}{2m}
        +\frac{\hbar k_{\text{SO}}}{m} \left(p_\gamma + q
        A_\gamma\right) \left(\hat{\gamma}\times \hat{z} \right)\cdot
        \vec{\sigma},
\end{equation}
%
%
where $m$ is the electron effective mass, $\vec{A}$ the vector
potential, and $k_{\text{SO}}$ the SO coupling strength.  The SO
coupling strength $k_{\text{SO}}$ is related to the spin precession
length $L_{\text{SO}}$ by $L_{\text{SO}} = \pi / k_{\text{SO}}$.  For
InAs quantum wells the spin-precession length ranges from $0.2$ to $1$
$\mu$m~\cite{nitta-1997, schaepers-1998,
grundler-2000,schaepers-2004,miller-2003}.  We neglect the Zeeman
splitting introduced by the magnetic field.  The wavefunction on a
bond (quantum wire) connecting two generic nodes $\alpha$ and $\beta$
along the direction $\hat{\gamma}_{\alpha\beta}$ reads
%
%
\begin{eqnarray}\label{wavefunction}
        \mathbf{\Psi}_{\alpha\beta}(r)=\displaystyle\frac{e^{-if_{\alpha
        r}} e^{-i(
        \hat{\gamma}_{\alpha\beta}  \times\hat{z})\cdot
          \vec{\sigma}~ k_{\text{SO}} r}}{\sin(k
        l_{\alpha\beta})} \left\{ \sin\left[ k
        (l_{\alpha\beta}-r)\right] \mathbf{\Psi}_{\alpha} \nonumber
        \right. \\ \left.
        + \sin(k r)     e^{i f_{\alpha\beta}}
        e^{i(\hat{\gamma}_{\alpha\beta}  \times\hat{z})\cdot
          \vec{\sigma}~ k_{\text{SO}} l_{\alpha\beta}}
        \mathbf{\Psi}_{\beta} \right\}.
\end{eqnarray}
%
%
where $k$ is related to the eigen energy by 
$\epsilon=\frac{\hbar^2}{2m} (k^2-k_{\text{SO}}^2)$, $r$
is the coordinate along the bond, and $l_{\alpha\beta}$ the length of
the bond\cite{note2}.  The magnetic field gives rise to the phase factors
%
%
\begin{equation}\label{phase-vector-potential}
        \exp\left\{-i f_{\alpha,r} \right\} = \exp\left\{-i
        \frac{2\pi}{\phi_0}\displaystyle\int_\alpha^r 
        \vec{A}\cdot \text{d}\vec{l} \right\},
\end{equation}
%
%
where $\phi_0=h/e$ is the flux quantum.  The spinors
$\mathbf{\Psi}_{\alpha}$ and $\mathbf{\Psi}_{\beta}$ are the values of
the wavefunction at the nodes $\alpha$ and $\beta$ respectively.  The
spin precession due to the Rashba effect is described by the
exponentials containing Pauli matrices in Eq.~(\ref{wavefunction}).

Equation~(\ref{wavefunction}) is the key step to generalize the
existing methods to study quantum networks
\cite{kottos-1999,vidal-tra-2000,bercioux-2004} in the presence of
Rashba SO coupling and magnetic field.  The wavefunction of the whole
network is obtained by imposing the conservation of probability current
at every node. For a generic node $\alpha$ it reads:
%
%
\begin{equation}\label{continuity}
        \textbf{M}_{\alpha\alpha} \mathbf{\Psi}_\alpha+\sum_{\langle
        \alpha,\beta\rangle}
        \textbf{M}_{\alpha\beta}\mathbf{\Psi}_\beta=0,
\end{equation}  
%
%
where
%
%
\begin{subequations} \label{ms}
\begin{eqnarray} 
        \textbf{M}_{\alpha\alpha} &= & \sum_{\langle
        \alpha,\beta\rangle} \cot k l_{\alpha\beta}\\
        \textbf{M}_{\alpha\beta} &= & -
        \displaystyle\frac{e^{
         -i f_{\alpha\beta}} 
        e^{i(
           \hat{\gamma}_{\alpha\beta}  \times\hat{z})\cdot
          \vec{\sigma}~
        k_{\text{SO}}l_{\alpha\beta}}
        }{ \sin k l_{\alpha\beta}}.
\end{eqnarray}
\end{subequations}
%
%
In Eqs.~(\ref{continuity}) and (\ref{ms}) the sum $\sum_{\langle
\alpha,\beta\rangle}$ runs over all nodes $\beta$ which are connected
by a bond to the node $\alpha$.  This set of boundary conditions
ensures the self-adjointness of the Schr\" odinger operator for the
whole network.

For a closed network, the secular equation is derived requiring a
non-trivial solution for the set of equations obtained writing the
condition Eq.~(\ref{continuity}) at every node.  A similar approach,
making use of the Bloch condition, can be used for computing the
spectrum of an infinite periodic network.

%
%
\begin{figure*}[!th]
        \begin{minipage}{0.45\textwidth} \centering
        \includegraphics[width=2.5in]{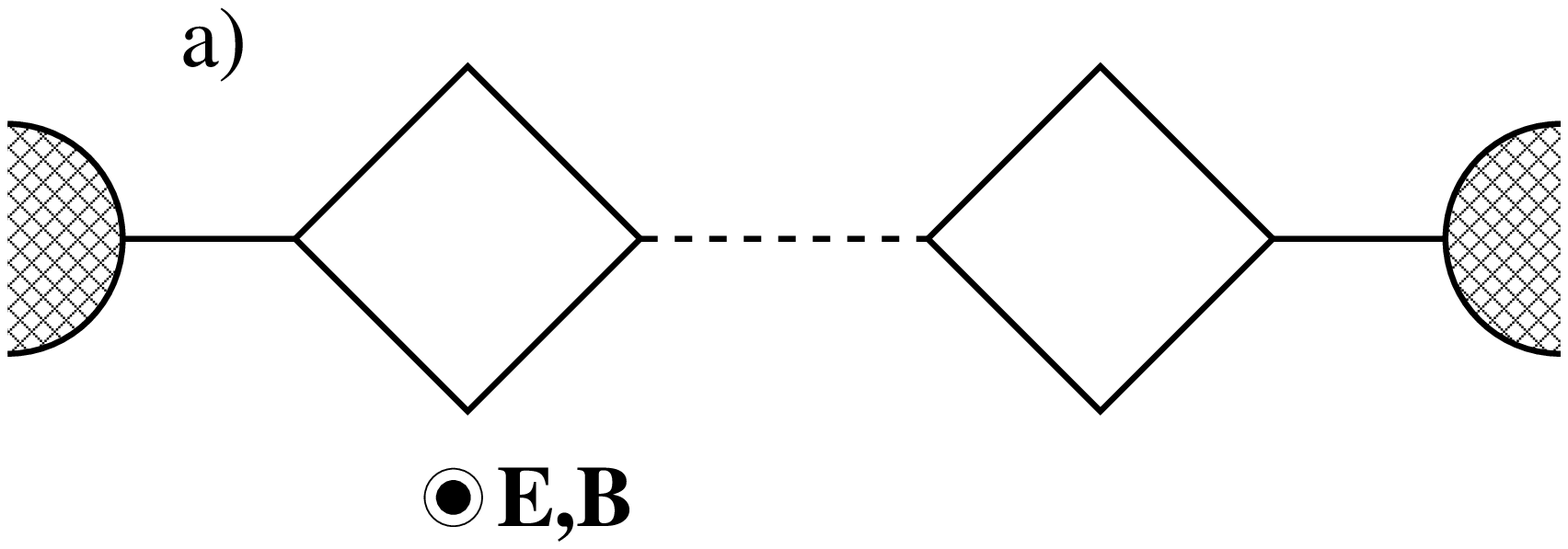} \end{minipage}
        \begin{minipage}{0.45\textwidth} \centering
        \includegraphics[width=3.in]{figure1b.eps} \end{minipage}
        \begin{minipage}{0.45\textwidth} \centering
        \includegraphics[width=3.in]{figure1c.eps} \end{minipage}
        \begin{minipage}{0.45\textwidth} \centering
        \includegraphics[width=3.in]{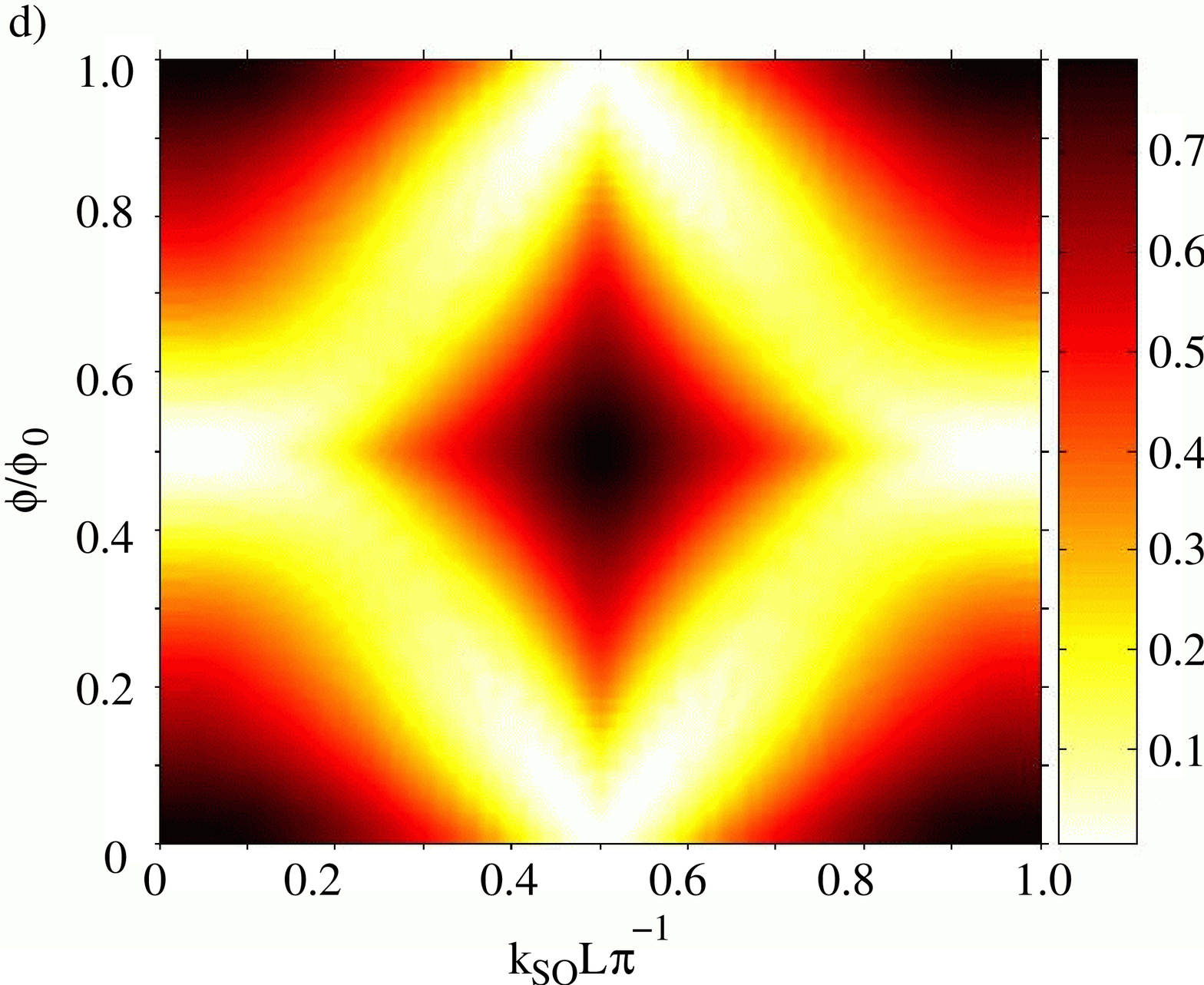} \end{minipage}
        \caption{(color online) Panel a): A finite-size piece of the diamond chain
        connected to reservoirs. 
        There are 1 input and 
        1 output single-channel leads. 
        The number of input channels, considering spin, is $N_{\text{in}}=2$.\\
        Panel b): Averaged conductance per
        channel, $\langle G\rangle_k / N_{\text{in}}$,
        as a function of the reduced flux (solid
        line) evaluated at $k_{\text{SO}}L\pi^{-1} = 0$, and of
	spin-orbit coupling evaluated at $\phi/\phi_0 =
	0$ (dashed line) for the diamond chain with 10 elementary squares.  \\      
        Panel c): Averaged conductance per
        channel, $\langle G\rangle_k / N_{\text{in}}$, as a function of the
        reduced flux evaluated at $k_{\text{SO}}L\pi^{-1} = 0.5$
        (solid line), and of spin-orbit coupling evaluated at
        $\phi/\phi_0 = 0.5$ (dashed line).\\ 
        Panel d): Color-scale plot of the averaged conductance 
        $\langle G\rangle_k$ as a function of the reduced flux and spin-orbit 
        coupling. \\ 
\label{diamondfig}}
\end{figure*}
%
%

In this paper, we are mainly concerned in calculating linear transport 
through a finite-size network connected to external leads. 
The linear conductance can be evaluated making use of the 
Landauer--B\"uttiker scattering formalism
\cite{landauer-1988,buttiker-1988}. We proceed along the lines
proposed by Vidal {\it et al.}~\cite{vidal-tra-2000}. Here, we
describe the method for the case of only two external leads; a
generalization to many leads is straightforward.  In the two
output/input semi-infinite leads, we assume that there is no SO
coupling, and no magnetic field.  To compute the transmission
coefficients we inject from the input wire an electron with spin
$\sigma=\pm$ along a generic direction, whose corresponding spinors
are $\mathbf{\chi}_{\sigma}$.  The wavefunctions on the external leads
are simply
%
%
\begin{eqnarray}
        \Psi_{\text{in}}(r) &=& e^{i k_{\text{in}} r}
        \mathbf{\chi}_\sigma + \sum_{\sigma^\prime}
        r_{\sigma^\prime\sigma} e^{-i k_{\text{in}} r}
        \mathbf{\chi}_{\sigma^\prime}\\ \Psi_{\text{out}}(r) &=&
        \sum_{\sigma^\prime} t_{\sigma^\prime\sigma} e^{i
        k_{\text{in}} r} \mathbf{\chi}_{\sigma^\prime},
\end{eqnarray} 
%
%
where $r$ is the coordinate on the semi-infinite input/output lead,
with the origin fixed at the position of the input/output node. The
transmission and reflection coefficients ($t_{\sigma^\prime\sigma}$
and $ r_{\sigma^\prime\sigma}$, respectively) can be obtained by
solving the linear system of equations arising from the continuity of
the probability current at all nodes in the network, and of the
wavefunction at the input and output nodes.  The conditions for the
continuity of the probability current at internal nodes are given in
Eq.~(\ref{continuity}). For the external nodes they read
%
%
\begin{eqnarray}
        \textbf{M}_{00} \mathbf{\Psi}_0 +\sum_{\langle 0 ,\beta
        \rangle} \textbf{M}_{0\beta}\mathbf{\Psi}_\beta =-
        i(\mathbf{\chi}_\sigma-\sum_{\sigma^\prime}
        r_{\sigma^\prime\sigma} \mathbf{\chi}_{\sigma^\prime})\\
        \textbf{M}_{NN} \mathbf{\Psi}_N +\sum_{\langle N ,\beta
        \rangle}\textbf{M}_{N\beta}\mathbf{\Psi}_\beta = i
        \sum_{\sigma^\prime} t_{\sigma^\prime\sigma}
        \mathbf{\chi}_{\sigma^\prime},
\end{eqnarray} 
%
%
where the injection node is labeled as ``$0$'' and the output node as
``$N$''.  The total transmission coefficient is then simply
$|t|^2=\sum_{\sigma,\sigma^\prime} |t_{\sigma^\prime\sigma}|^2$.

We now introduce some concepts useful for understanding the
interference phenomena due to magnetic field and Rashba SO coupling.
Let us suppose that an electron is at node $\alpha$ with a spinor
$\mathbf{\Psi}_\alpha$ and that it travels along the bond from
$\alpha$ to $\beta$. When it reaches node $\beta$ it has acquired a
phase.  Its spinor in $\beta$ is related to the one in $\alpha$ by
%
%
\begin{equation}\label{phases}
        \mathbf\Psi_{\beta}= R_{\text{dyn}}(l_{\alpha\beta},k)
        R_{\text{AB}}(\alpha,\beta) \mathbf{R_{\text{SO}}} (\alpha,
        \beta) \mathbf\Psi_{\alpha},
\end{equation}
%
%
The dynamical phase is simply $R_{\text{dyn}}(l_{\alpha\beta},k)=
\exp(i k l_{\alpha\beta})$. 
The AB phase is 
%
%
\begin{equation}
        R_{\text{AB}}(\alpha,\beta) = \exp\left\{-i
        \displaystyle\frac{2 \pi}{\phi_0}
        \displaystyle\int_{\alpha}^{\beta} \vec{A}\cdot
        \text{d}\vec{l} \right\}.
\end{equation}
%
%
The SO coupling, on the other hand, introduces a non-Abelian phase (a
rotation in spin space), which reads
%
%
\begin{equation}\label{non:abelian}
        \mathbf{R_{\text{SO}}}(\alpha,\beta)=\exp\left[-i(
        \hat{\gamma}_{\alpha\beta} \times\hat{z})\cdot \vec{\sigma}~
        k_{\text{SO}} l_{\alpha\beta}\right].
\end{equation}
%
%
Equation~(\ref{phases}) can be used to understand the difference
between the interference induced by an AB phase and a non-Abelian
phase. Let us consider a rhombus-like loop. An electron is injected at
one of the vertices and collected at the opposite one. The phase
(\ref{phases}) acquired by the electron wavefunction is different
along the two possible paths. Destructive interference is achieved
only when the two partial waves for these paths sum to zero at the
final point. A straightforward calculation shows that it is possible
to obtain destructive interference induced by the AB phase in every
rhombus-like loop. Instead, the non-Abelian phase
Eq.~(\ref{non:abelian}), gives rise to destructive interference only
for the special case of a square.

%
%
\begin{figure*}[t]
        \begin{minipage}{0.45\textwidth} \centering
        \includegraphics[width=2.5in]{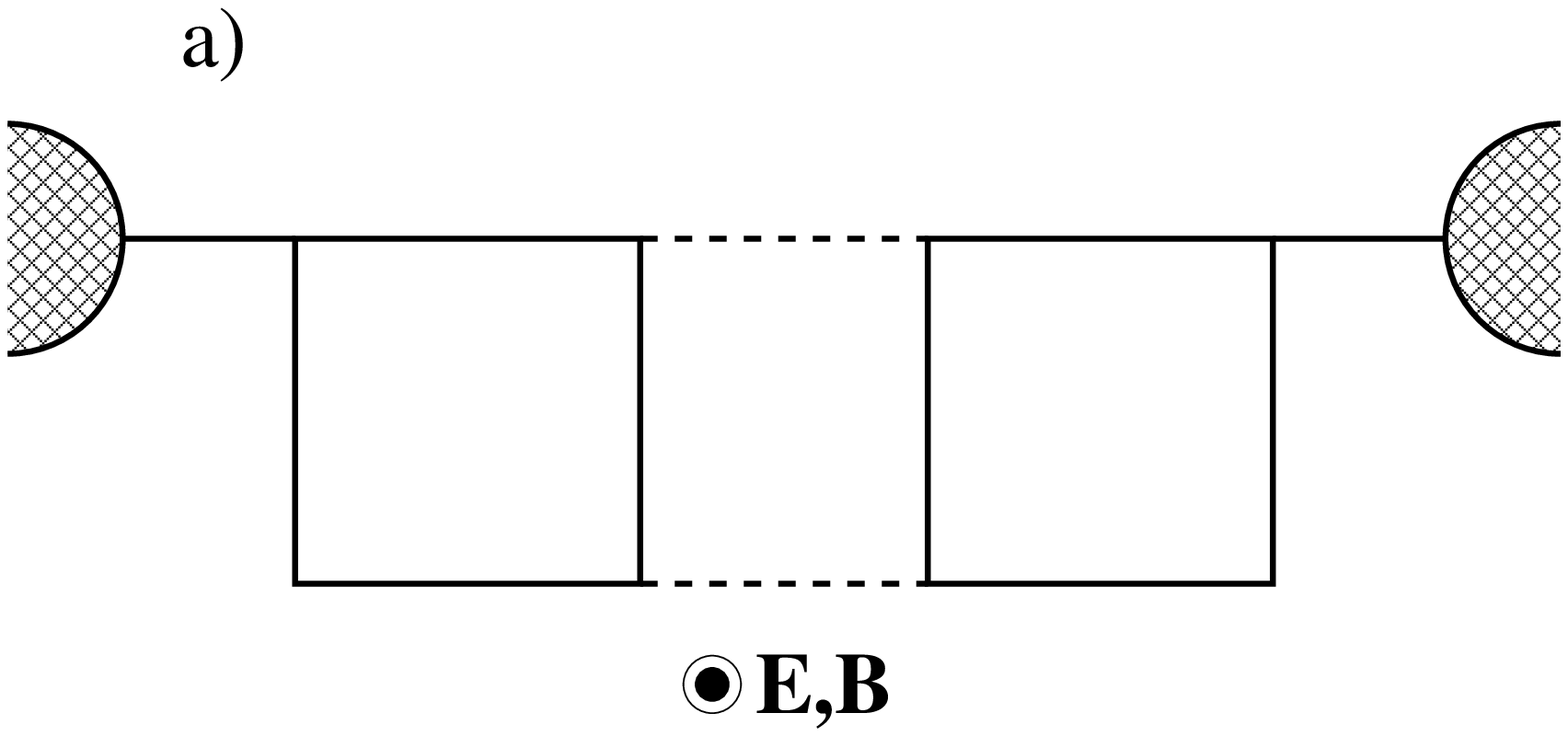} \end{minipage}
        \begin{minipage}{0.45\textwidth} \centering
        \includegraphics[width=3.in]{figure2b.eps} \end{minipage}
        \begin{minipage}{0.45\textwidth} \centering
        \includegraphics[width=3.in]{figure2c.eps}
        \end{minipage} \begin{minipage}{0.45\textwidth} \centering
        \includegraphics[width=3.in]{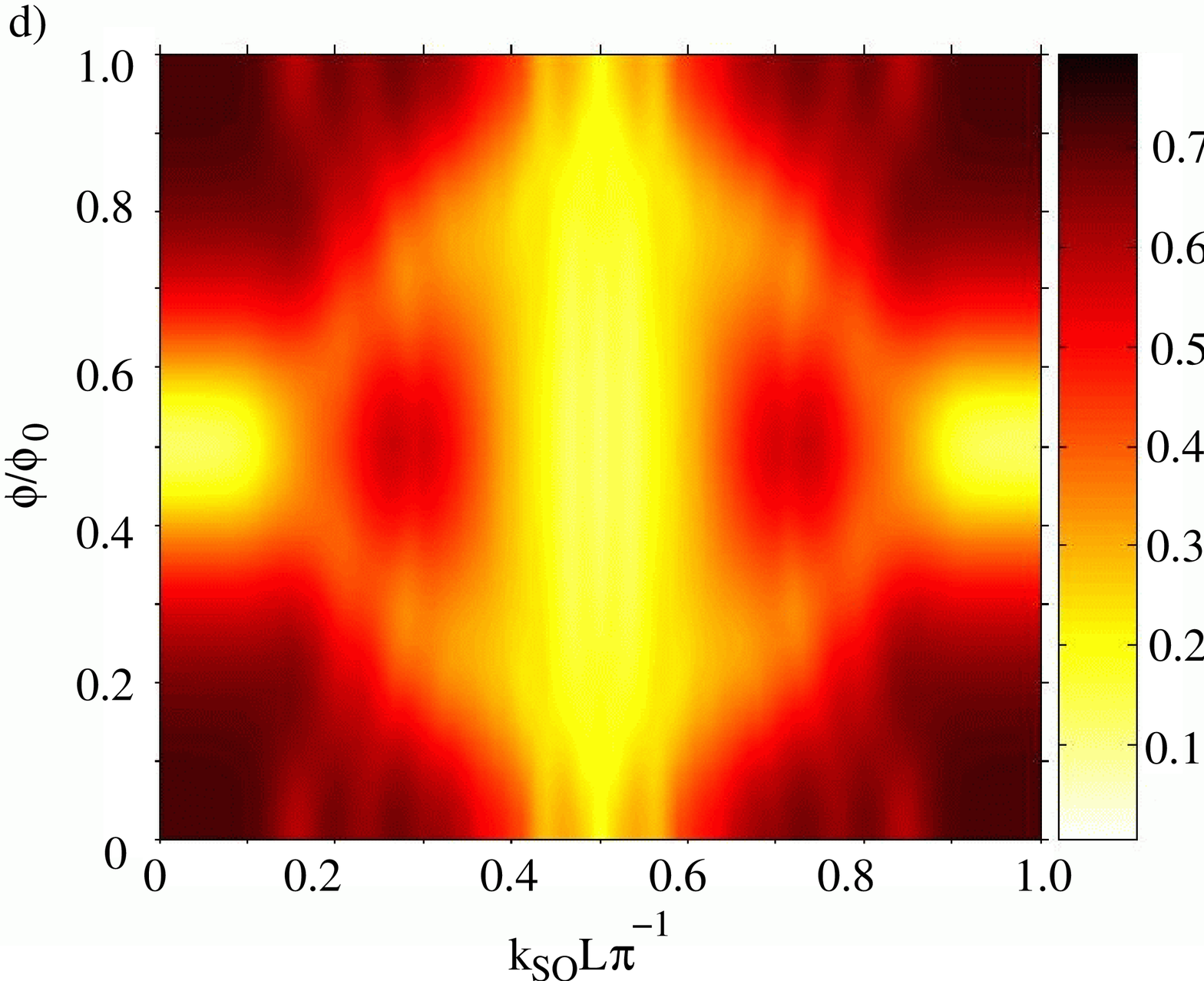} \end{minipage}
        \caption{(color online)
        Panel a): A finite-size piece of the square ladder
        connected to reservoirs. 
        There are 1 input and 
        1 output single-channel leads.
        The number of input channels, considering spin, is $N_{\text{in}}=2$.\\  
        Panel b): Averaged conductance per
        channel, $\langle G\rangle_k / N_{\text{in}}$,
        as function of the reduced flux (solid
        line), and of spin-orbit coupling (dashed line) for the
        ladder with 10 elementary squares.  \\       
        Panel c): Averaged conductance per
        channel, $\langle G\rangle_k / N_{\text{in}}$, as function of the reduced
        flux evaluated at $k_{\text{SO}}L\pi^{-1} = 0.5$ (solid line), 
        and of spin-orbit coupling evaluated at $\phi/\phi_0 = 0.5$ (dashed line).\\
        Panel d): Color-scale plot of the averaged conductance 
        $\langle G\rangle_k$ as a function of the reduced flux and spin-orbit 
        coupling.\\ 
        \label{ladderfig}}
\end{figure*}
%
%
\begin{figure}
        \includegraphics[width=3.in]{figure3a.eps}\\
        \includegraphics[width=3.in]{figure3b.eps} 
\caption{Panel a):   
        Disorder-averaged conductance per channel, $\langle
        G\rangle_{dis} / N_{\text{in}}$, as function of the reduced flux for zero
        spin-orbit coupling (solid line), and of spin-orbit coupling
        for zero magnetic field (dashed line), for the
        diamond chain with 10 elementary squares.\\  
        Panel b):       
        Disorder-averaged conductance per channel, $\langle
        G\rangle_{dis} / N_{\text{in}}$, 
        as function of the reduced flux for zero
        spin-orbit coupling (solid curve), and of spin-orbit coupling
        for zero magnetic field (dashed curve), for the
        ladder with 10 elementary squares.\\ 
        In both panels, the injection
        wave-vector is uniformly distributed in
        $[k_{\text{F}}-\frac{\pi}{2 L}, k_{\text{F}}+\frac{\pi}{2 L}]$
        with $k_{\text{F}} L=100$, and the disorder strength is
        $\Delta L/L = 0.01$ for the main panel, and $\Delta L/L =
        0.05$ for the inset.
\label{disorder1d}}
\end{figure}
%
%
\begin{figure*}[t]
        \begin{minipage}{0.45\textwidth} \centering
        \includegraphics[width=3.in]{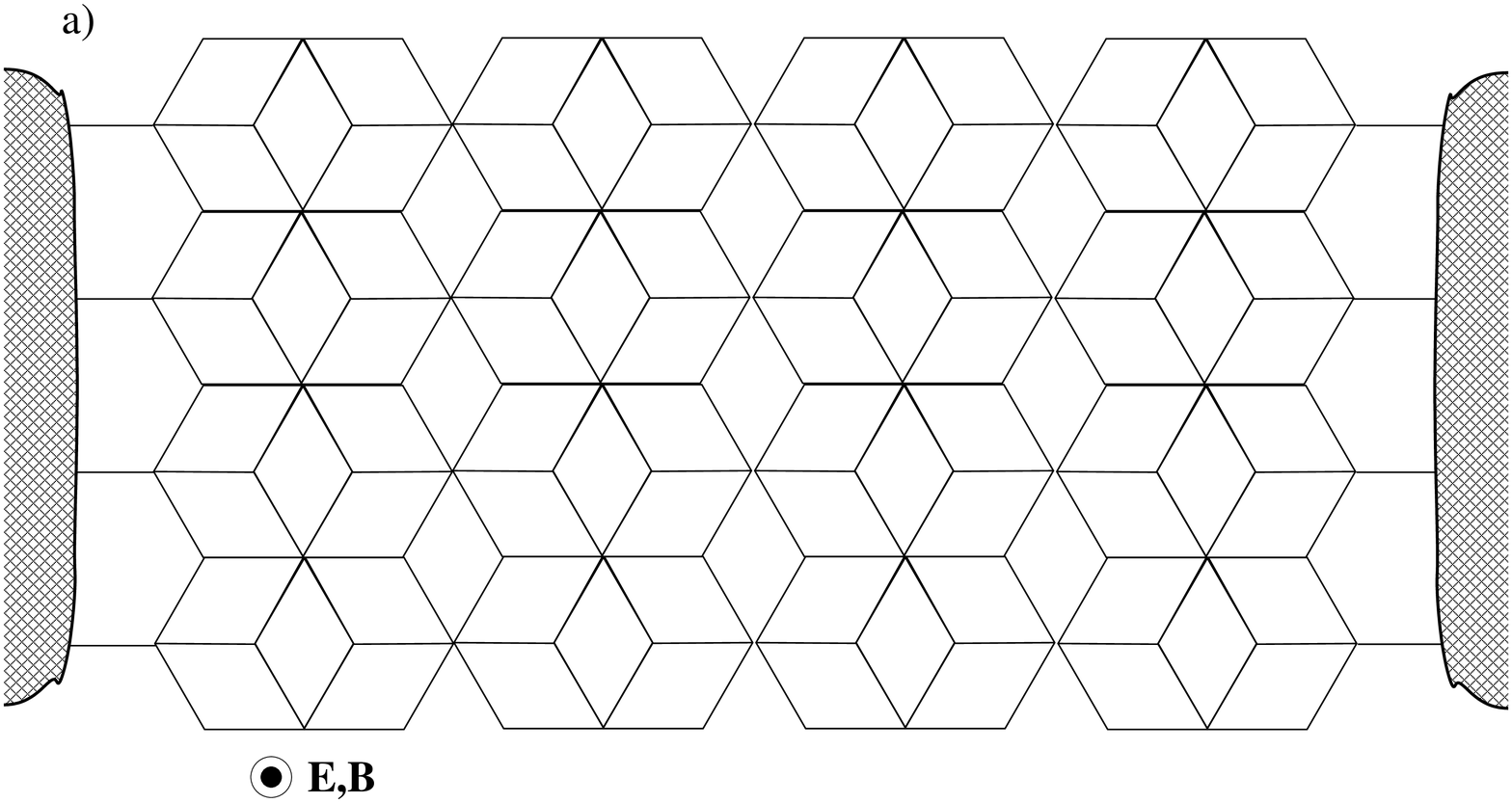} \end{minipage}
        \begin{minipage}{0.45\textwidth} \centering
        \includegraphics[width=3.in]{figure4b.eps} \end{minipage}
        \begin{minipage}{0.45\textwidth} \centering
        \includegraphics[width=3.in]{figure4c.eps} \end{minipage}
        \begin{minipage}{0.45\textwidth} \centering
        \includegraphics[width=3.in]{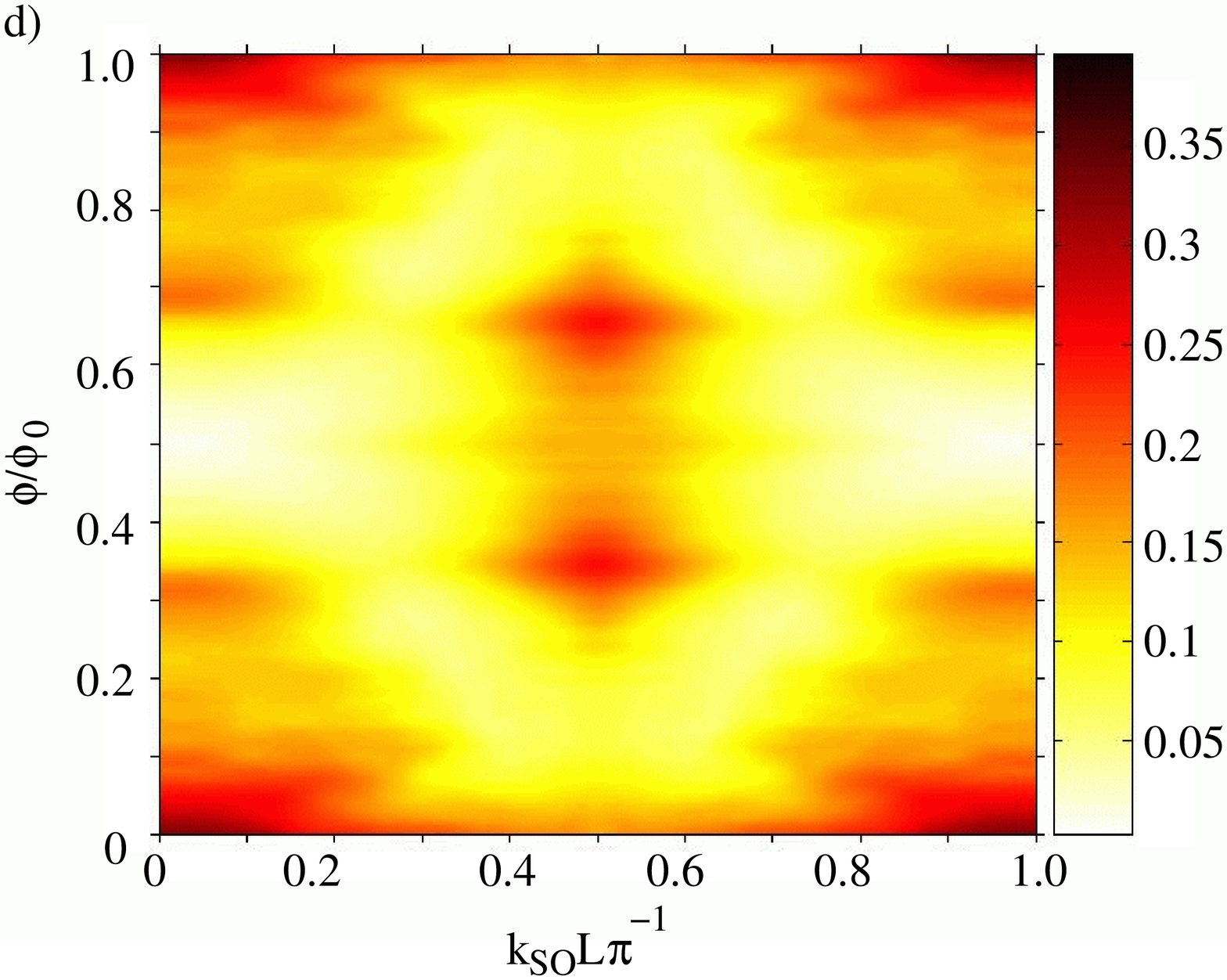} \end{minipage}
        \caption{(color online)
        Panel a): A finite-size piece of the $\T_3$ network
        connected to reservoirs.   
        There are 4 input and 
        4 output single-channel leads. 
        The number of input channels, considering spin, is $N_{\text{in}}=8$.\\
        Panel b): Averaged conductance
        per channel, $\langle G\rangle_k / N_{\text{in}}$, as a function of
        the reduced flux evaluated at $k_{\text{SO}}L\pi^{-1} = 0$
	(solid line), and of spin-orbit coupling evaluated at $\phi/\phi_0 = 0$
        (dashed line) for the $\T_3$ lattice with 200 quantum wires
        (89 rhombi).\\ Panel c): Averaged conductance per
        channel, $\langle G\rangle_k / N_{\text{in}}$, as a function of the
        reduced flux evaluated at $k_{\text{SO}}L\pi^{-1} = 0.5$
        (solid line), and of spin-orbit coupling evaluated at
        $\phi/\phi_0 = 0.5$ (dashed line).\\ 
        Panel d): Color-scale plot of the      
        averaged conductance per channel, $\langle G\rangle_k/N_{\text{in}}$, as a function 
        of the reduced 
        flux and spin-orbit coupling.
\label{tau_3}}
\end{figure*}
%
%
\begin{figure*}[t]
        \begin{minipage}{0.45\textwidth} \centering
        \includegraphics[width=3.in]{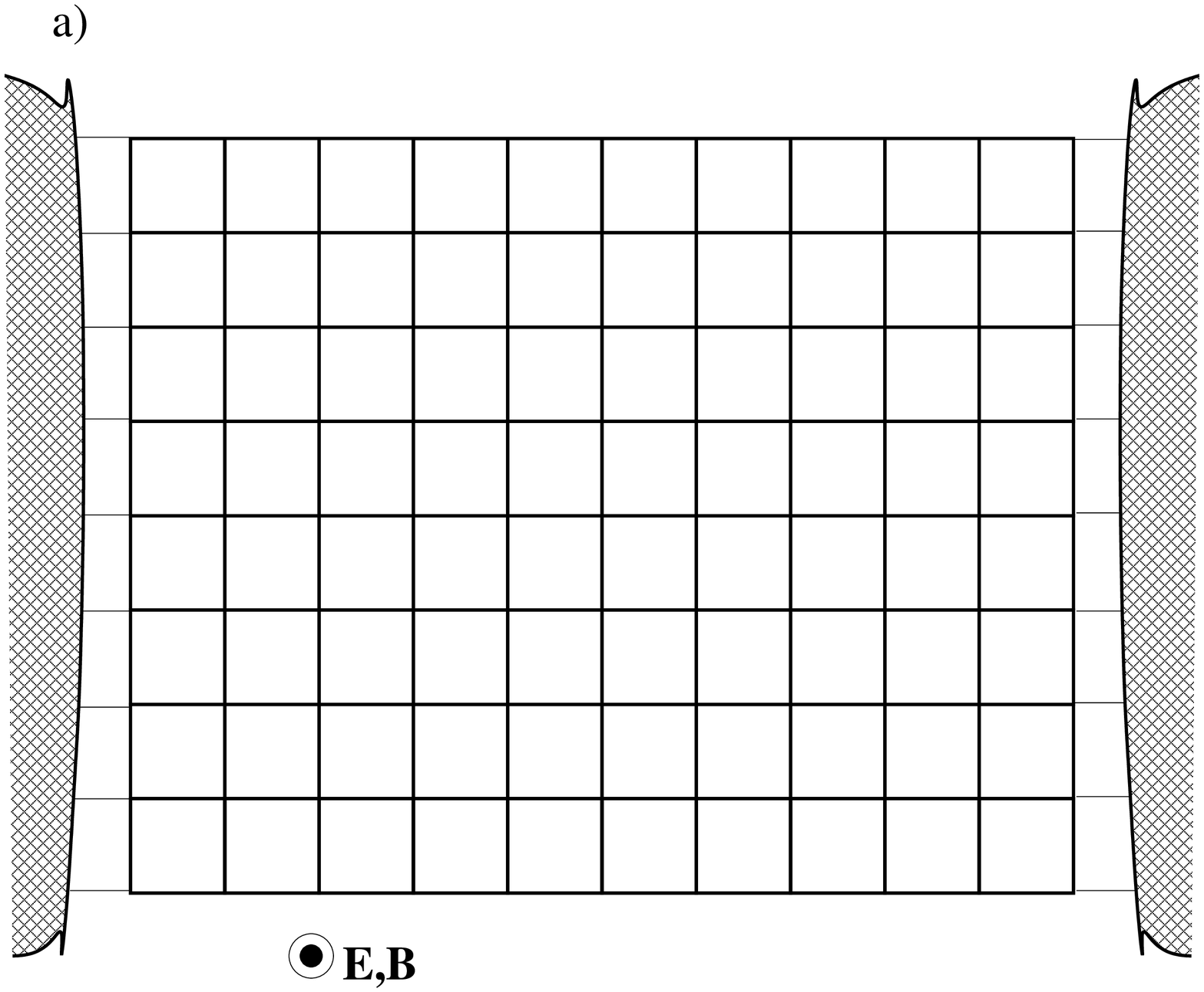} \end{minipage}
        \begin{minipage}{0.45\textwidth} \centering
        \includegraphics[width=3.in]{figure5b.eps} \end{minipage}
        \begin{minipage}{0.45\textwidth} \centering
        \includegraphics[width=3.in]{figure5c.eps}
        \end{minipage} \begin{minipage}{0.45\textwidth} \centering
        \includegraphics[width=3.in]{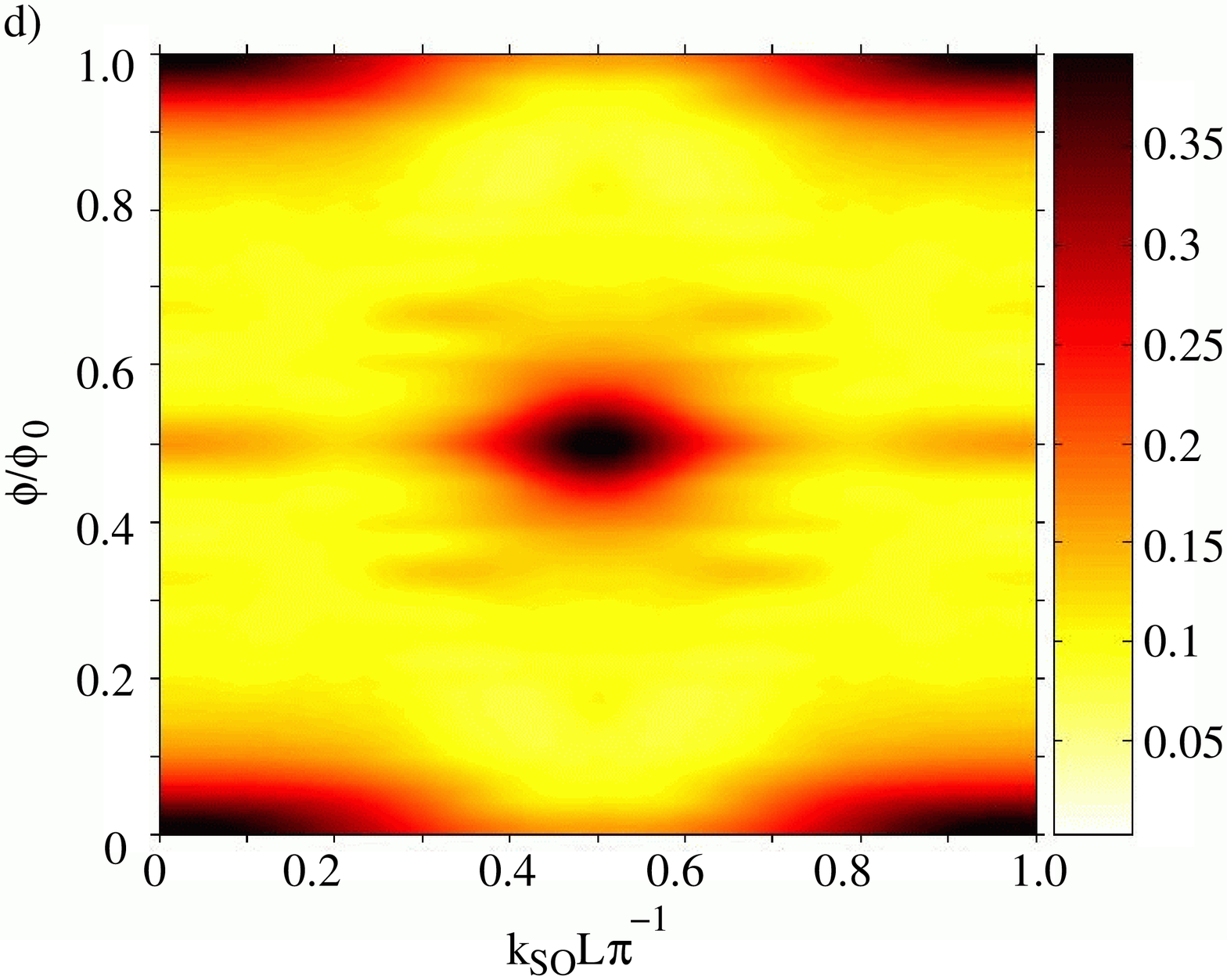} \end{minipage}
        \caption{(color online) 
        Panel a): A finite-size piece of the square lattice
        connected to reservoirs.   
        There are 9 input and 
        9 output single-channel leads. The number of input channels, considering spin, 
        is $N_{\text{in}}=18$.\\ 
        Panel b): Averaged conductance
        per channel, $\langle G\rangle_k / N_{\text{in}}$, as a function of
        the reduced flux evaluated at $k_{\text{SO}}L\pi^{-1} = 0$
	(solid line), and of spin-orbit coupling evaluated at $\phi/\phi_0 = 0$
        (dashed line) for the square lattice with 178 quantum wires
        (80 squares).  \\ Panel c): Averaged conductance per
        channel, $\langle G\rangle_k / N_{\text{in}}$, as a function of the
        reduced flux evaluated at $k_{\text{SO}}L\pi^{-1} = 0.5$
        (solid line), and of spin-orbit coupling evaluated at
        $\phi/\phi_0 = 0.5$ (dashed line).\\ 
        Panel d): Color-scale plot of the      
        averaged conductance per channel, $\langle G\rangle_k/N_{\text{in}}$, as a function 
        of the reduced 
        flux and spin-orbit coupling.\label{square}}      
\end{figure*}
%
%
\begin{figure}
        \includegraphics[width=3.in]{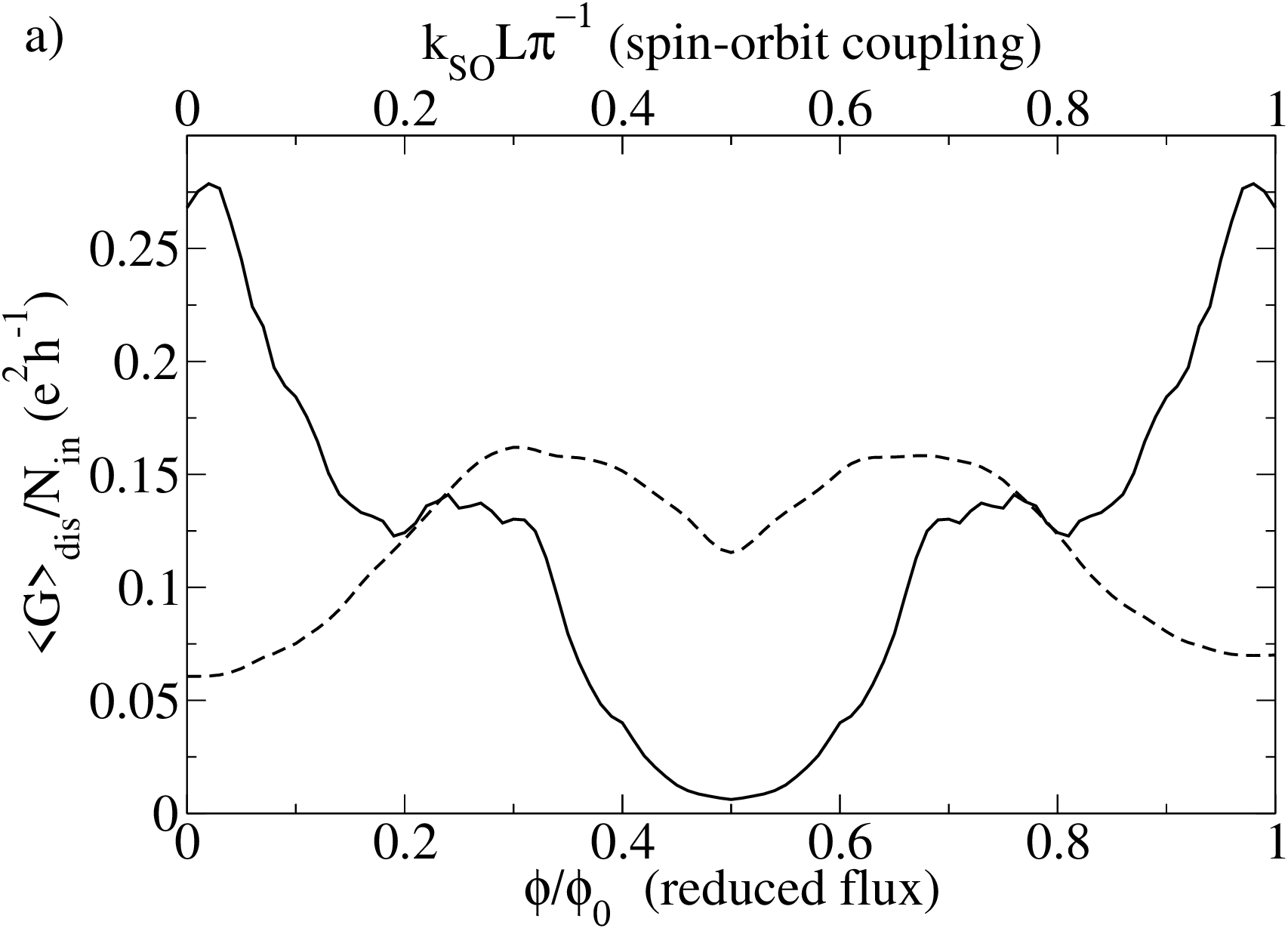}\\
        \includegraphics[width=3.in]{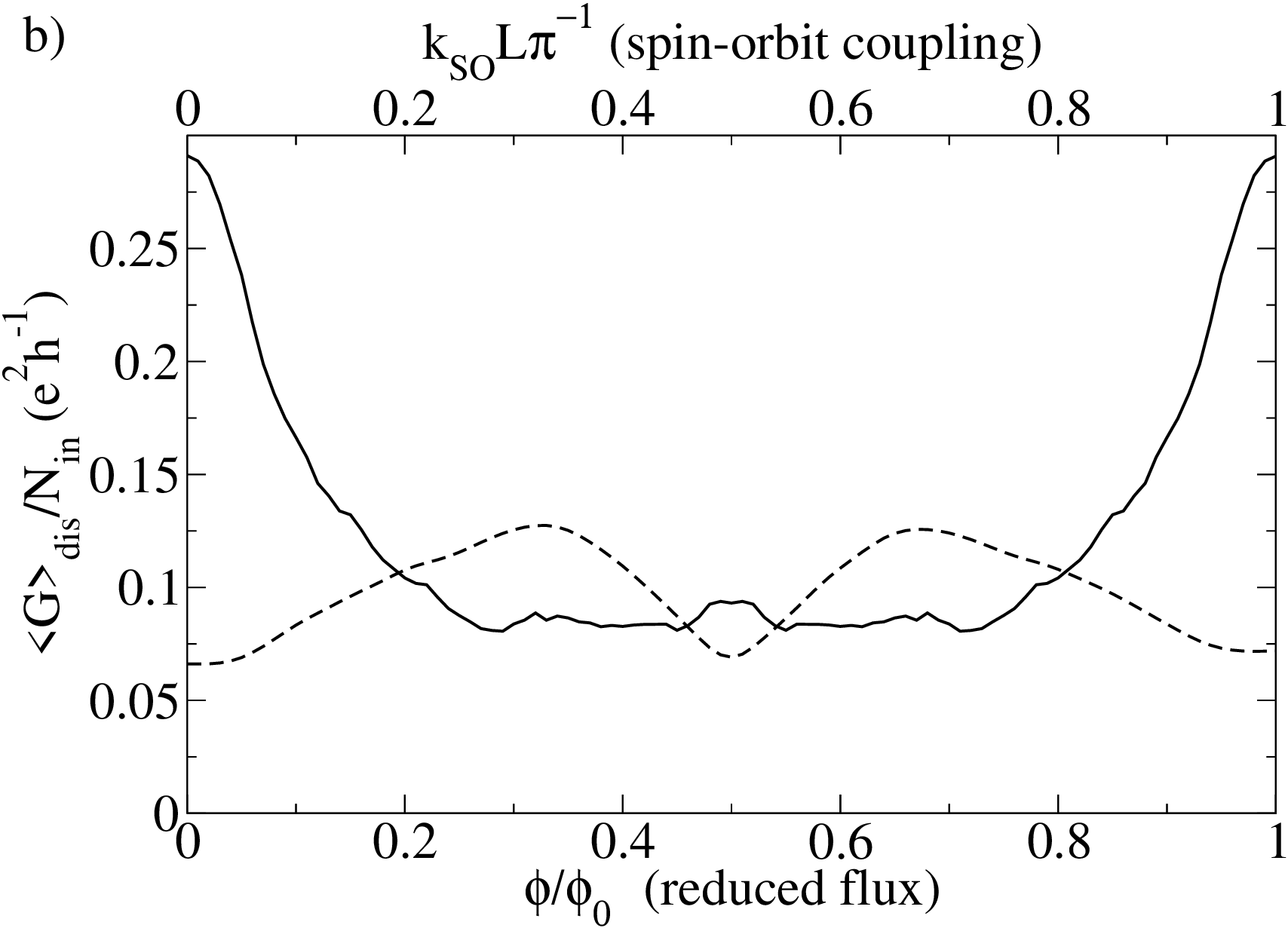} 
\caption{
        Panel a) Disorder-averaged conductance per channel, $\langle
        G\rangle_{dis} / N_{\text{in}}$, as a function of the reduced flux
        (solid line), and of spin-orbit coupling (dashed line), 
        for the $\T_3$ lattice with 200 quantum wires (89 rhombi). \\ 
        Panel b) Disorder-averaged conductance per channel, $\langle
        G\rangle_{dis} / N_{\text{in}}$, as a function of the reduced flux
        (solid line) and of spin-orbit coupling (dashed line), 
         for the square lattice with 178 quantum wires
        (80 squares).\\     
        In both panels, the
        injection wave-vector is uniformly distributed in
        $[k_{\text{F}}-\frac{\pi}{2 L}, k_{\text{F}}+\frac{\pi}{2 L}]$
        with $k_{\text{F}} L=100$, and the disorder strength is
        $\Delta L/L = 0.05$.\label{disorder2d}}
\end{figure}
\section{One-dimensional networks}\label{one-dim}

In this section we study linear transport through one-dimensional
networks. In particular we consider a \emph{diamond chain} [shown in
panel a) of Fig.~\ref{diamondfig}], i.e a chain of squares connected
at one vertex, and a square ladder [shown in panel a) of
Fig.~\ref{ladderfig}].

The diamond chain is a bipartite structure containing nodes with
different coordination numbers. For the diamond chain, electron
localization due to SO coupling has been demonstrated in
Ref.~\onlinecite{bercioux-2004}.  Here, we study the interplay of SO
coupling and of an orbital magnetic field.  The conductance is a
function of the injection wave-vector $k_{\text{in}}$, and it is
periodic in $k_{\text{in}}$ with period $\pi/L$, being $L$ the length
of a bond.  Furthermore, the conductance as a function of
$k_{\text{in}}$ shows a rich interference pattern. As we are not
interested in the interference due to the dynamical phases
$R_{\text{dyn}}$, we have made the usual choice [see
Ref.~\onlinecite{vidal-tra-2000}] to integrate the conductance over
$k_{\text{in}}\in[0, \pi/L]$. Throughout this paper, the integrated
conductance will be indicated by $\langle G \rangle_{k}$.  We note
that finite temperature or finite voltage will introduce in a natural
way an average over $k_{\text{in}}$. For $\mbox{Max}[{K_{\text{B}} T,
eV}] \ge K_{\text{B}} T^{*}= \frac{\hbar^2}{ m} k_F\frac {\pi}{L}$,
$\langle G\rangle_{k}$ will be the quantity measured in the transport
experiments.  Taking for the Fermi energy of the single-channel wires
$10$ meV, $m/m_e=0.042$ for the effective mass (InAs), and $L=1\mu$m,
yields $T^{*}\approx 7$ K.

In panel b) of Fig.~\ref{diamondfig}, we show the conductance $\langle
G\rangle_{k}$ as a function of the reduced flux for zero SO coupling
(solid line), and as a function of the SO-coupling strength for zero
magnetic field (dashed line). The reduced flux is defined as the
number of flux quanta per unit rhombus.  In this geometry, when either
magnetic field or SO coupling is present, the integrated conductance
reaches zero both as a function of the reduced flux (zero SO coupling)
and as a function of SO coupling strength (zero magnetic field). In
the first case the conductance vanishes when the reduced flux is a
half integer, in the latter when $k_{\text{SO}}L/\pi$ is a half
integer.  One can easily prove, making use of Eq.~(\ref{phases}), that
the vanishing conductance is due to the full interference of the
partial waves traveling on the upper and lower arm of a diamond. In
such a situation electron localization is achieved.  Depending on the
mechanism which leads to localization, this effect is called AB-cage
effect or Rashba-cage effect.
  
It is interesting to study the combined effect of the magnetic field
and of the SO coupling. This is done in panel c) of
Fig.~\ref{diamondfig}, where the conductance $\langle G\rangle_{k}$ is
plotted as a function of the SO coupling for $\phi/\phi_0 = 0.5$
(dashed line), and as function of the magnetic field for
$k_{\text{SO}}L\pi^{-1} = 0.5$ (solid line).  Let us concentrate on
the fixed-SO-coupling case (solid line): when the reduced flux is 0 or
1, we are in a condition of complete interference induced by the SO
coupling; when the flux is moved away from these values, the AB phase
changes and the interference is no more fully destructive. This give
rise to the typical anti-localization peak shown in panel c).  A very
similar analysis applies to the fixed-flux case (dashed line).  The
full dependence of the conductance $\langle G\rangle_{k}$ on both the
reduced flux and SO coupling is shown, as a color-scale plot, in panel
d) of Fig.~\ref{diamondfig}. The anti-localization peak is clearly
visible in the center of the plot.

Disorder is inevitable in nanostructures and its effect needs to be
accounted for. From previous studies, both the Rashba-cage effect and
the AB-cage effect are expected to be robust against disorder.  We
consider a model where the length of each bond is randomly distributed
in the interval $[L-\Delta L,L+\Delta L]$, and we average observable
quantities over disordered configurations.  The half width of the
distribution $\Delta L$ gives the strength of the disorder.  This type
of disorder is particularly dangerous because it affects the phases
acquired by electrons when traveling along the bonds.  
We consider the physical situation $\Delta L / L\ll 1$. Under this
condition the dispersion of loop areas is negligible, and the
periodicity with reduced flux is preserved. Consistently, we neglect
the effect of the bond-length fluctuations on the AB
phases~\cite{vidal-tra-2000}.

In the absence of disorder, the conductance is periodic in the flux
$\phi/\phi_0$ and in $k_{\text{SO}} L/\pi$ with periodicity $1$.
Increasing disorder the periodicity with magnetic field and spin-orbit
will be eventually halved.  The halving of the oscillation period is
due to the Altshuler-Aronov-Spivak (AAS) effect~\cite{aas-1981}, and
it is related to the enhancement of back-reflection due to
interference of pair of paths traveling clockwise and
counter-clockwise along a square of the chain (according to weak
localization picture).  A detailed study of the period-halving
transition as a function of disorder strength for the $\T_3$ lattice
with magnetic field, has been presented in
Ref.~\onlinecite{vidal-tra-2000}. Our findings agree with the scenario
presented there.

The disordered-average conductance for the diamond chain is shown in
panel a) of Fig.~\ref{disorder1d}.  For moderate-strength disorder
($k_{\text{F}} \Delta L\approx 1$) we find both the AB- and the
Rashba-cage effect are preserved. Only at higher disorder strengths
[see the inset of panel a)] the halving of the periodicity takes
place.

The square ladder has a complete different topology and this is
reflected in its transport properties. In particular, no localization
is expected to occur. In Fig.~\ref{ladderfig} and in panel b) of
Fig.\ref{disorder1d}, for the sake of comparison, the same quantities
displayed previously for the diamond chain, are shown for the square
ladder. The main results being that no localization occurs, and that
in the presence of disorder AAS oscillation are present.


\section{Two-dimensional systems}\label{twodim}

We now turn our attention to two-dimensional networks.  In particular,
we consider transport through a finite piece of $\T_3$ lattice and
contrast it with transport through a finite piece of square lattice.
The way the finite-size networks are connected to reservoirs, kept at
different chemical potentials, is shown in panel a) of
Figs.~\ref{tau_3} and \ref{square}.

The $\T_3$ lattice [shown in panel a) of Fig.~\ref{tau_3}] is a
periodic hexagonal structure with three sites per unit cell, one
sixfold coordinated and two threefold coordinated. This is an example
of two-dimensional regular bipartite lattice containing nodes with
different coordination numbers .

We show that in this kind of structure transport properties exhibit a
signature of the interference effects due to Rashba SO coupling and to
magnetic field.  In fact we expect complete localization by means of
magnetic field~\cite{vidal-tra-2000}, but not of Rashba SO coupling,
as can be predicted by considering interference of partial waves with
the phase factors given in Eq.~(\ref{phases}).

In panel b) of Fig.~\ref{tau_3}, we show the conductance $\langle
G\rangle_{k}$ through a finite piece of $\T_3$ lattice, as a function
of the reduced flux for zero SO coupling (solid line), and as a
function of the SO-coupling strength for zero magnetic field (dashed
line).  When the SO coupling is absent we observe a suppression of the
conductance at half-integer values of the reduced flux, due to the
existence of the AB-cage effect.  The residual value of the
conductance minimum is not zero due to the existence of dispersive
edge states~\cite{note1}. This residual value is independent of the
number of injection channels.  On the other hand, when the magnetic
field is absent we do not observe such a strong suppression of the
conductance as a function of the SO coupling. A minimum is still
present, but this is due to partial interference, which does not
induce complete localization. Furthermore, the finite conductance at
the minimum cannot be attributed to the existence of edge states,
because its value depends on the number of injection channels.

In panel c) of Fig.~\ref{tau_3}, we show the conductance $\langle
G\rangle_{k}$ as a function of the SO coupling for $\phi/\phi_0 = 0.5$
(dashed line), and as a function of the magnetic field with
$k_{\text{SO}}L\pi^{-1} = 0.5$ (solid line).  In the case of fixed
finite SO coupling, the behavior of the conductance as a function of
reduced flux is qualitatively similar to when the SO coupling is
absent.  In particular, a well defined minimum for $\phi/\phi_0=0.5$
is still observed.  On the other hand, for fixed magnetic field, the
SO coupling suppresses the destructive interference due to the AB
effect and an anti-localization peak takes place.  Panel d) of
Fig.~\ref{tau_3} shows the full dependence of the conductance $\langle
G\rangle_{k}$ as a function of both reduced flux and SO coupling.

In panel a) of Fig.~\ref{disorder2d} we show the behavior of the
disorder-averaged conductance as a function of the reduced flux (solid
line) for zero SO coupling, and as function of the SO coupling (dashed
line) for zero magnetic field.  The periodicity of the
disorder-averaged conductance as function of $k_\text{SO} L/\pi$ is no
longer $1$ but $1/2$. The oscillation with period 1 have been washed
out, while those with period $1/2$ are still present since they are
related to phase-coherent pairs of time reversed trajectories
according to the weak-localization pictures.  This is consistent with
the fact that in the $\T_3$ the SO coupling does not induce complete
localization, and therefore, the oscillation with period $1$ are not
protected against disorder.  On the other hand, the disorder-averaged
conductance as a function of the reduced flux remains, for this
disorder strength, still $\phi_0$-periodic.

We now consider transport through a finite-size square lattice [shown
in panel a) of Fig.~\ref{square}].  This network, unlike the $\T_3$
lattice, does not fulfill the necessary condition to exhibit the AB-
or Rashba-cage effect, i.e.  it does not present a bipartite structure
containing nodes with different coordination numbers.  Accordingly, we
do not expect any electron localization phenomenon caused either by
the magnetic field or the SO coupling.

In panel b) of Fig.\ref{square}, we show the conductance $\langle
G\rangle_{k}$ through a finite-size square lattice as a function of
the reduced flux for zero SO coupling (solid line), and as a function
of the SO-coupling strength for zero magnetic field (dashed line).
The overall behavior of the conductance as a function of magnetic
field (zero SO coupling) and of SO coupling (zero magnetic field) is
very different.  However it has to be noticed that both curves reach
the same value, respectively, at $\phi/\phi_0=1/2$ (solid line) and
$k_{\text{SO}}L\pi^{-1}=1/2$ (dashed line).  It is interesting to
analyze what happens when both the magnetic field and the SO coupling
are present.  In panel c) of Fig.~\ref{tau_3}, we show the conductance
$\langle G\rangle_{k}$ as a function of the SO coupling for
$\phi/\phi_0 = 0.5$ (dashed line), and as a function of the magnetic
field with $k_{\text{SO}}L\pi^{-1} = 0.5$ (solid line).  The behavior
of both curves is very similar.  In particular, the conductance shows
an anti-localization-like peak both as a function of SO coupling and
of magnetic field, respectively, around $\phi/\phi_0 = 0.5$ and
$k_{\text{SO}}L\pi^{-1} = 0.5$.  The full dependence of the
conductance $\langle G\rangle_{k}$ as a function of both reduced flux
and SO coupling is shown in panel d) of Fig.~\ref{square}, where it
can be seen that significative conductance is obtained only in the
anti-localization peak at the center of the plot.

In panel b) of Fig.~\ref{disorder2d}, we show the disorder-averaged
conductance as a function of the reduced flux (solid line) for zero SO
coupling, and as function of the SO coupling (dashed line) for zero
magnetic field.  In this case, we notice that the oscillations of
period $1$ as a function of magnetic field are more robust than those
as a function of SO coupling (but not as robust as in the $\T_3$ when
localization is achieved).

\section{Conclusion}
We have introduced a formalism to study quantum networks made by
single-channel quantum wires in the presence of Rashba spin-orbit
coupling and of magnetic field.  In particular, we have investigated
the interplay of the AB phases and of the non-Abelian phases
introduced by SO coupling in transport through finite-size
one-dimensional and two dimensional networks.

While SO coupling can induce localization in particular
one-dimensional networks\cite{bercioux-2004}, complete localization by
means of SO coupling has not been found in the two-dimensional $\T_3$
lattice.  However, signatures of the SO coupling are still visible in
the transport properties, and intriguing effects occur due to the
presence of both magnetic field and SO coupling.

\begin{acknowledgments}
Fruitful discussions with C. Cacciapuoti, G. De Filippis, P. Lucignano
and C.A. Perroni (\textit{Federico II} University, Naples, Italy)
and with R. Fazio, D. Frustaglia and M. Rizzi (\textit{Scuola Normale
Superiore}, Pisa, Italy) are gratefully acknowledged. Finally DB
wishes to acknowledge A. Cer\'e (\textit{Hippos Campi Flegrei}, 
Naples, Italy).
\end{acknowledgments}

%
%


\begin{thebibliography}{99}
%
\bibitem{vidal-1998} J. Vidal, R. Mosseri, and B. Dou\c cot,
Phys. Rev. Lett. \textbf{81}, 5888 (1998).
%
\bibitem{bohm} Y. Aharonov, and D. Bohm Phys. Rev. \textbf{115}, 485
(1959). 
%
\bibitem{anderson-1958} P.W. Anderson, Phys. Rev. \textbf{109}, 1492
(1958). 
%
\bibitem{vidal-prb-long}  J. Vidal, P. Butaud, B. Dou\c cot, and
R. Mosseri, Phys. Rev. B \textbf{64}, 155306 (2001)
%
\bibitem{vidal-cage-2000} J. Vidal, B. Dou\c cot, R. Mosseri, and
P. Butaud, Phys. Rev. Lett. \textbf{85}, 3906 (2000).
%
\bibitem{vidal-tra-2000} J. Vidal, G. Montambaux, and B. Dou\c cot,
Phys. Rev. B  \textbf{62}, R16294 (2000).
%
\bibitem{korshunov} S. E. Korshunov, Phys. Rev. B \textbf{63}, 134503
(2001).
%
\bibitem{cataudella} V. Cataudella and R. Fazio,
Europhys. Lett. \textbf{61}, 341 (2003). 
%
\bibitem{abilio-1999} C.C. Abilio, P. Butaud, T. Fournier,
B. Pannetier, J. Vidal, S. Tedesco, and B. Dalzotto,
Phys. Rev. Lett. \textbf{83}, 5102 (1999).
%
\bibitem{naud-2001} C. Naud, G. Faini, and D. Mailly,
Phys. Rev. Lett. \textbf{86}, 5104 (2001).
%
\bibitem{casher} Y. Aharonov and A. Casher,
Phys. Rev. Lett. \textbf{53}, 319 (1984).
%
\bibitem{mathur} H. Mathur, and A. D. Stone, Phys. Rev. Lett. 
{\textbf 68}, 2964 (1992)
%
\bibitem{balatsky} A.V. Balatsky, and B. L. Altshuler, Phys. Rev. Lett.  
\textbf{70} 1678 (1993).
%
\bibitem{aronov-1993} A.~G. Aronov and Y.~B. Lyanda-Geller,
Phys. Rev. Lett. \textbf{70}, 343 (1993).
%
\bibitem{splettstoesser-2003} J. Splettst\"o{\ss}er, M. Governale,
and U. Z\" ulicke, Phys. Rev. B \textbf{68}, 165341 (2003).
%
\bibitem{frustaglia} D. Frustaglia and K. Richter, Phys. Rev. B
\textbf{69}, 235310 (2004).
%
\bibitem{rashba-1960} E.I. Rashba, Fiz. Tverd. Tela (Leningrad)
\textbf{2}, 1224 (1960), [Sov. Phys. Solid State 2, 1109 (1960)].
%
\bibitem{rashba-1984} Y.A. Bychkov and E.I. Rashba, J. Phys. C
\textbf{17}, 6039 (1984).
%
\bibitem{nitta-1997} J. Nitta, T. Akazaki, H. Takayanagi and T. Enoki,
Phys. Rev. Lett. \textbf{78}, 1335 (1997).
%
\bibitem{schaepers-1998} T. Sch\" apers, J. Engels, T. Klocke,
M. Hollfelder and H. L\" uth, J. Appl. Phys.\textbf{83}, 4324 (1998).
%
\bibitem{grundler-2000} D. Grundler, Phys. Rev. Lett. \textbf{84},
6074 (2000).
%
\bibitem{schaepers-2004} T. Sch\" apers, J. Knobbe, and V.A. Guzenko,
Phys. Rev. B \textbf{69}, 235323 (2004).
%
\bibitem{miller-2003} J.B. Miller, D.M. Zumb\" uhl, C.M. Marcus,
Y.B. Lyanda-Geller, D. Goldhaber-Gordon, K. Campman, and A.C. Gossard,
Phys. Rev. Lett. \textbf{90}, 76807 (2003).
%
\bibitem{bercioux-2004} D. Bercioux, M. Governale, V.
Cataudella, and V. M. Ramaglia,
Phys. Rev. Lett. \textbf{93}, 56802 (2004).
%
\bibitem{note2}The term
in $k_{\text{SO}}^2$ can be neglected in realistic situations.
%
\bibitem{kottos-1999}
T. Kottos {and} U. Smilansky, Ann. Phys. (N.Y.) \textbf{274}, 76 (1999).
%
\bibitem{landauer-1988}
R. Landauer, IBM J. Res. Dev. \textbf{1}, 223 (1957). 
%
\bibitem{buttiker-1988} M. Buttiker, IBM J. Res. Dev. \textbf{32},
 317 (1988).
%
\bibitem{aas-1981} B. Altshuler, A. G.  Aronov and B. Spivak, Pis'ma
Zh. Eksp. Teor. Fiz. \textbf{33}, 101 (1981), [JEPT Lett.
  \textbf{33}, 94, (1981)].
%
\bibitem{note1} An analysis of edge-state formation in the $\T_3$
lattice can be found in Ref.~\onlinecite{vidal-prb-long}. 
%
\end{thebibliography}
\end{document}